\documentclass[twocolumn]{aastex631}
\usepackage{amsmath}
\usepackage{esint}
\usepackage{ragged2e}
\usepackage{hyperref}
\usepackage{physics}
\usepackage{bm}

\defcitealias{Menou2012b}{M12}
\defcitealias{Hardy2022}{H22}
\defcitealias{Hardy2023}{H23}

\newcommand{\Rm}{{\rm Rm}}

\shorttitle{Hot spot offset from thermoresistive instability in hot jupiters}

\begin{document}

\title{Hot Spot Offset Variability from Magnetohydrodynamical Thermoresistive Instability in Hot Jupiters}

\author[0000-0002-2599-6225]{Rapha\"el Hardy}
\affiliation{D\'epartement de Physique, Universit\'e de Montr\'eal, Montr\'eal, QC, H3C 3J7, Canada\\}
\affiliation{Department of Physics and Trottier Space Institute, McGill University, Montr\'eal, QC, H3A 2T8, Canada\\}
\affiliation{Institut Trottier de Recherche sur les Exoplan\`etes (iREx), Universit\'e de Montr\'eal, Montr\'eal, QC H3C 3J7, Canada}

\author[0000-0003-1618-3924]{Paul Charbonneau}
\affiliation{D\'epartement de Physique, Universit\'e de Montr\'eal, Montr\'eal, QC, H3C 3J7, Canada\\}

\author[0000-0002-6335-0169]{Andrew Cumming}
\affiliation{Department of Physics and Trottier Space Institute, McGill University, Montr\'eal, QC, H3A 2T8, Canada\\}
\affiliation{Institut Trottier de Recherche sur les Exoplan\`etes (iREx), Universit\'e de Montr\'eal, Montr\'eal, QC H3C 3J7, Canada}




\begin{abstract}

Hot Jupiter atmospheres are possibly subject to a thermoresistive instability. Such an instability may develop as the ohmic heating increases the electrical conductivity in a positive feedback loop, which ultimately leads to a runaway of the atmospheric temperature. We extend our previous axisymmetric one-dimensional radial model, by representing the temperature and magnetic diffusivity as a first order Fourier expansion in longitude. This allows us to predict the hot spot offset during the unfolding of the thermoresistive instability and following Alfv\'enic oscillations. We show a representative simulation undergoing the thermoresistive instability, in which the peak flux offset varies between approximately $\pm 60^{\circ}$ on timescales of a few days with potentially observable brightness variations. Therefore, this thermoresistive instability could be an observable feature of hot Jupiters, given the right timing of observation and transit and the right planetary parameters.

\end{abstract}



\section{Introduction} \label{sec:intro}
    
    Hot Jupiter (HJ) atmospheres are interesting case studies of extreme atmospheric dynamics.
    Being tidally locked due to their proximity ($\lesssim 0.1 ~{\rm AU}$) to their host star, the extreme radiative heating on their dayside sustains large temperature gradients between their locked day and nightside, which drives equatorial jets \citep{Showman2011,Komacek2016,Read2018,Imamura2020}. These jets are usually prograde in both observations and hydrodynamical simulations \citep{Showman2002,Cooper2005,Showman2009,Rauscher2010,Kataria2016}. The jets advect the hot spot, the hottest region of the atmosphere, away from the substellar point either eastward or westward depending on the direction of the zonal winds. There are, however, a few exceptions displaying retrograde jets \citep{Armstrong2016,Dang2018,Bell2019,Jackson2019,vonEssen2020}. 

    The temperature regime characterizing HJs leads to partial ionization of their atmospheres, which can then couple to the magnetic field \citep{Rogers2014b}. Magnetic effects have been proposed as the cause of retrograde winds and westward hot spots, as well as atmospheric variability \citep{Rogers2014b,Rogers2017b,Hindle2019,Hindle2021b,Hindle2021a,Hardy2022,Hardy2023}. The main source of ionization is alkali metals such as potassium and sodium, which have low first ionization energies \citep{Perna2010a,Batygin2010}. Previous studies have investigated magnetic coupling between the winds in the upper atmosphere and the magnetic field pervading the planetary interior, leading to slower prograde equatorial jets, or even retrograde jets \citep{Rauscher2013,Rogers2017b,Menou2012b,Hardy2022,Hardy2023}.

    As the alkali metals just start being ionized at the temperatures of HJ atmospheres, the ionization fraction is very temperature sensitive. Thus, small temperature variations can have very large impact on the magnetic diffusivity (MD) $\eta$ and magnetic coupling \citep{Perna2010a}. \cite{Menou2012b} showed that with $\eta$ decreasing as temperature increases can drive a thermoresistive instability (TRI), leading to runaway ohmic heating of the atmosphere (see also \citealt{Hubbard2012} and \citealt{Price2012}). \cite{Rauscher2013} included this effect in 3D atmospheric circulation models under the assumption that the magnetic Reynolds number Rm remains small and so the magnetic forces can be treated as a drag term \citep{Perna2010a}. 
    
    In order to explore the full range of dynamics including the regime of large Rm where the plasma is strongly-coupled to the field, we developed a simplified axisymmetric model of the equatorial plane of a HJ (\citealt{Hardy2022,Hardy2023}, hereafter \citetalias{Hardy2022} and \citetalias{Hardy2023}). The model assumes that angular momentum is continuously pumped into the equatorial plane from higher latitudes, driving a zonal flow at the equator \citep{Showman2011}. The radial component of the magnetic field at the equator supports torsional Alfven waves and interacts with the flow. This geometry enables a study of the feedback between ohmic heating, the evolving $\eta$, and the dynamics, while also including the full radial structure of the atmosphere. We showed that the outcome of the TRI in HJs with intermediate temperatures (equilibrium temperatures $T_\mathrm{eq}\approx 1000$--$1200\ \mathrm{K}$) is to create bursts of Alfven oscillations separated by longer periods of quiescence. These time-dependent bursting solutions are not present when $\eta$ is assumed to be time-independent; they represent a new class of time-dependent behaviour driven by the temperature-dependence of $\eta$. 


    In this paper, we extend the model of \citetalias{Hardy2023} by relaxing the assumption of axisymmetry. We do this by using a low order Fourier expansion in the azimuthal angle $\phi$, following the approach of \cite{Tritton1988} who studied the problem of convection in a torus. This expansion enables us to include the variation in $\eta$ with $\phi$ in an approximate way, allowing a study of the displacement angle of the hot spot (temperature maximum) during the different phases of the bursts created by the TRI.
    
    The paper is organized as follows. We describe the model in Section \ref{sec:model}, and discuss in detail the properties of a single a representative solution, followed by Section \ref{sec:results} where a briefer discussion of behavior variations across the model's parameter space.  
    We close the paper (Section \ref{sec:discussion}) by discussing possible observational signatures of the TRI, including variations in the hot spot offset and the atmospheric temperature.
    
\section{Extended One Dimensional Model in the Equatorial Plane}\label{sec:model}

    \subsection{Model Setup}

        Following \citetalias{Hardy2023}, we consider the atmospheric layer extending from a pressure of $1.0$~bar at the base to $0.01$~bar at the top, with gravitational acceleration $g_p=9.0 \rm ~m~s^{-2}$. We assume that the atmosphere is in hydrostatic balance and composed of an ideal gas of pure molecular hydrogen. As the modeled layer represents only 3\% of the planetary radius, we adopt the plane parallel approximation where we map the the spherical coordinates $(\phi,\theta,r)$ (longitude, latitude and radius, respectively) onto Cartesian coordinates $(x,y,z)$. 
        We assume that the flow velocity is in the longitudinal direction and depends only on height, i.e. $u_x(z)$. incompressibility ($\nabla \cdot {\bf u} = 0$) then implies $u_y=u_z=0$. The magnetic field, satisfying $\nabla \cdot {\bf B} = 0$, is
        \begin{equation}\label{eq:magnetic_field}
            {\bf B}(z,t)=B_x(z,t){\hat x}+B_0{\hat z},
        \end{equation}
        where $B_0$ is a background radial field and $B_x$ is the toroidal field induced by differential rotation. As discussed by \citetalias{Hardy2023}, such a radial field component could arise from a misaligned dipole, higher order multipole, or a local dynamo (e.g.~\citealt{Rogers2017a,Dietrich2022}). The simple magnetic field geometry described by Equation~(\ref{eq:magnetic_field}) can still support  torsional Alfv\'en oscillations impacting zonal flows.

        \citetalias{Hardy2023} assumed full axisymmetry. We relax this here by allowing the temperature $T$ to depend on $x$ (corresponding to a longitudinal variation in $T$). Note that $u_x$ and $B_x$ remain axisymmetric: they must be independent of $x$ in order for the velocity and magnetic field to remain divergence free.
        With these assumptions, the magnetohydrodynamics (MHD) equations \citep{Davidson2001} reduce to 
%
        \begin{equation}\label{eq:movement}
            \frac{\partial u_x}{\partial t} = \frac{B_0}{{\mu}_0 {\bar \rho}}  \frac{\partial B_x}{\partial z}  + \frac{{\bar \mu}}{{\bar \rho}}  \frac{\partial^2 u_x}{\partial z^2}  + a_{x},
        \end{equation}
        %
        \begin{equation}\label{eq:induction}
            \frac{\partial B_x}{\partial t} = B_0  \frac{\partial u_x}{\partial z}   +  \frac{\partial \eta}{\partial z}  \frac{\partial B_x}{\partial z}  + \eta  \frac{\partial^2 B_x}{\partial z^2},
        \end{equation}
        %
        \begin{equation}\label{eq:temperature}
        \begin{aligned}
                \frac{\partial T}{\partial t} + u_x \frac{\partial T}{\partial x} &=  \frac{1}{{\bar \rho} c_p}  \frac{\partial \bar \chi}{\partial z}  \frac{\partial T}{\partial z}  + \frac{\bar \chi}{{\bar \rho} c_p}  \frac{\partial^2 T}{\partial z^2}  + \frac{1}{{\bar \rho} c_p} \frac{\partial F_{\rm irr}}{\partial z}\\ &+ \frac{{\bar \mu}}{{\bar \rho} c_p} \left( \frac{\partial u_x}{\partial z} \right)^2 + \frac{\eta}{{\mu}_0 {\bar \rho} c_p}  \left( \frac{\partial B_x}{\partial z} \right) ^2.
        \end{aligned}
        \end{equation}
        These are the same set of equations considered by \citetalias{Hardy2023} except for the advective term on the left hand side of Equation (\ref{eq:temperature}), which allows for the longitudinal advection of temperature. Since the layer is thin, we assume that other terms involving horizontal gradients, for example in the ohmic dissipation term, are negligible compared to vertical gradients. We use an overbar to indicate quantities that are not time-evolved (see Section \ref{sec:BCs&ICs} for further details).

        Equation (\ref{eq:movement}) is the $x$-component of the incompressible Navier-Stokes equation where $\bar \rho$ is the gas density, and $\bar \mu$ is the dynamic viscosity. The acceleration term
        \begin{equation}\label{eq:velocity_forcing}
            a_{x} = \dot{v} \exp({-P/\mathrm{bar}})~,
        \end{equation}
        models the effects of angular momentum transfer to the equator from higher latitudes, with the pressure $P$ measured in bars, and $\dot{v}$ setting the peak amplitude. Equation (\ref{eq:induction}) is the $x$-component of the induction equation. The temperature-dependent MD is given by 
        %
        \begin{equation}\label{eq:diffusivity}
        \eta(T) = 0.023 \frac{\sqrt{T}}{\chi_e} {\rm m^2~s^{-1}},
        \end{equation}
        taken from \citet{Menou2012a} and based on the results of \citet{Draine1983}. The ionization fraction $\chi_e$ is obtained from the Saha equation \citep{Rogers2014b} adopting solar abundances as given in \citet{Lodders2010} considering sodium and potassium only (these elements give the dominant contribution to the ionization).
        The heat capacity is $c_p$, and the thermal conductivity is defined as
        \begin{equation}\label{eq:thermal_conductivity}
            {\bar \chi} = \frac{16 \sigma T^3}{3\kappa_{\rm th}{\bar \rho}},
        \end{equation}
        where $\sigma$ is the Stefan-Boltzmann constant and $\kappa_{\rm th}$ is the Rosseland mean opacity, operating mostly in the infrared regime, where thermal emissions takes place. We use the thermal conductivity to set the dynamic viscosity
        \begin{equation}
            {\bar \mu} = \frac{\Pr \bar \chi}{c_p}.
        \end{equation}
        where the Prandtl number $\mathrm{Pr}$ is assumed constant throughout the atmosphere.

        In Equation (\ref{eq:temperature}) which gives the temperature evolution, we write the irradiation flux as 
        \begin{equation}\label{eq:flux_irradiation}
            F_{\rm irr} = F_s(\phi) \exp(-\sqrt{3}\kappa_v P/g),
        \end{equation}
        where $F_s(\phi)$ is the incoming flux from the host star at the surface of the planet set by the irradiation temperature, $\kappa_v$ is the visible opacity which we keep constant at  $4.0~\times~10^{-4}~\rm m^2~kg^{-1}$ throughout all our simulations as in \citetalias{Menou2012b}, and the $\sqrt{3}$ factor comes from the exponential in Equation (29) of \citet{Guillot2010}.
        
    \subsection{Longitudinal Expansion}
        We now develop a low order expansion in the longitudinal direction, applying the approach\footnote{\cite{Tritton1988} studied convection in a vertical torus heated at the bottom and cooled at the top. Interestingly, for that case an expansion of $T(\phi)$ in $\cos$ and $\sin$ terms leads to a set of equations equivalent to the famous Lorenz equations \citep{Lorenz1963}.} of \citet{Tritton1988}. 
        We expand the $\phi$-dependence of temperature and MD as
        \begin{equation}\label{eq:temperature_expansion}
            \tilde{T}(\phi) = T_0 + T_1 \cos \phi + T_2 \sin \phi,
        \end{equation}
        \begin{equation}\label{eq:eta_expansion}
            \tilde{\eta}(\phi) = {\eta_0} + {\eta_1} \cos \phi + {\eta_2} \sin \phi.
        \end{equation}
        We also assume for simplicity that the angular variation of the irradiation flux is given by $F_s(\phi)=F_s(0) (1+\cos\phi)/2$, which has a maximum at the substellar point ($\phi=0$) and drops to zero at the anti-substellar point\footnote{We note that including a more realistic irradiation profile, $F_s(\phi)\propto \cos\phi$ for $|\phi|<\pi/2$ and $0$ for $\pi/2<|\phi|<\pi$ gives a similar low order expansion $F_s(\phi) = (F_s(0)/\pi)(1 + (\pi/4)\cos\phi)$.} ($\phi=\pi$).      
        Inserting these expansions into Equation (\ref{eq:temperature}), with $\phi = x/R$, and identifying terms that are independent of $\phi$, proportional to $\cos\phi$, or proportional to $\sin\phi$, gives evolution equations for the amplitudes $T_0$, $T_1$, and $T_2$: 
        \begin{eqnarray}\label{eq:temperature_0}
                \frac{\partial T_0}{\partial t} &=&  \frac{1}{{\bar \rho} c_p}  \frac{\partial \bar \chi}{\partial z}  \frac{\partial T_0}{\partial z}  + \frac{\bar \chi}{{\bar \rho} c_p}  \frac{\partial^2 T_0}{\partial z^2}  + \frac{1}{2}\frac{1}{{\bar \rho} c_p} \frac{\partial F_{\rm irr}}{\partial z}\nonumber\\ && + \frac{{\bar \mu}}{{\bar \rho} c_p} \left( \frac{\partial u_x}{\partial z} \right)^2 + \frac{\eta_0}{{\mu}_0 {\bar \rho} c_p}  \left( \frac{\partial B_x}{\partial z} \right) ^2,\\        
    \label{eq:temperature_1}
                \frac{\partial T_1}{\partial t} &=& -\frac{u_x T_2}{R} + \frac{1}{{\bar \rho} c_p}  \frac{\partial \bar \chi}{\partial z}  \frac{\partial T_1}{\partial z}  + \frac{\bar \chi}{{\bar \rho} c_p}  \frac{\partial^2 T_1}{\partial z^2}\nonumber\\ && + \frac{1}{2}\frac{1}{{\bar \rho} c_p} \frac{\partial F_{\rm irr}}{\partial z} + \frac{\eta_1}{{\mu}_0 {\bar \rho} c_p}  \left( \frac{\partial B_x}{\partial z} \right) ^2,\\
        \label{eq:temperature_2}
                \frac{\partial T_2}{\partial t} &=& \frac{u_x T_1}{R} + \frac{1}{{\bar \rho} c_p}  \frac{\partial \bar \chi}{\partial z}  \frac{\partial T_2}{\partial z}  + \frac{\bar \chi}{{\bar \rho} c_p}  \frac{\partial^2 T_2}{\partial z^2}\nonumber\\ && + \frac{\eta_2}{{\mu}_0 {\bar \rho} c_p}  \left( \frac{\partial B_x}{\partial z} \right) ^2.
        \end{eqnarray}

        The original 2D problem given by Equations (\ref{eq:movement})--(\ref{eq:temperature}) has been reduced to the set of coupled 1D equations (\ref{eq:movement}), (\ref{eq:induction}), and (\ref{eq:temperature_0})--(\ref{eq:temperature_2}). These can now be solved to find the time-evolution of $u_x(z,t)$, $B_x(z,t)$, and the temperature components $T_0(z,t)$, $T_1(z,t)$ and $T_2(z,t)$. One complication is that $\eta$ appears in the induction equation (Equation~(\ref{eq:induction})), introducing a $\phi$-dependent term into that equation. To remain consistent with our assumption of axisymmetric $B_x$, we use the axisymmetric term $\eta_0$ when evaluating those terms in the induction equation.

        To obtain the coefficients $\eta_0$, $\eta_1$, and $\eta_2$ at each time step, we first evaluate $\eta(\phi)$ using Equations (\ref{eq:diffusivity}) and (\ref{eq:temperature_expansion}). A least squares fit of the functional form of Equation (\ref{eq:eta_expansion}) to $\eta(\phi)$ then gives
            \begin{equation}
                \eta_0 = \frac{1}{n_\phi} \sum_\phi \eta(\phi),
            \end{equation}
            \begin{equation}
                \eta_1 = \frac{\sum_{\phi} \eta(\phi) \cos(\phi)}{\sum_{\phi} \cos^2(\phi)},
            \end{equation}
            \begin{equation}
                \eta_2 = \frac{\sum_{\phi} \eta(\phi) \sin(\phi)}{\sum_{\phi} \sin^2(\phi)},
            \end{equation}
        where the sums are over the grid of $n_\phi$ $\phi$-values at which $\eta(\phi)$ has been evaluated. Note that since $\eta$ is exponential in temperature, a sinusoidal expansion of $\eta(\phi)$ as given by Equation (\ref{eq:eta_expansion}) is not necessarily a good approximation. We find that this approach reproduces $\eta(\phi)$ to within a factor of 6 at worst, but within ten percent during most of the simulation, which is reasonable considering that the amplitude of $\eta$ can change by orders of magnitude during an oscillation (see Appendix \ref{appendix:diffusivity} for further details).

    \subsection{Boundary Conditions, Initial Conditions, and Choice of Parameter Values}\label{sec:BCs&ICs}
        The boundary conditions for $u_x$ and $B_x$ are the same as specified in \citetalias{Hardy2023}. We impose the inner layer of our domain to be corotating with the core of the planet, setting $u_x = 0$, and we set the outermost layer to be free of any mechanical stresses, setting $\partial u_x/\partial z = 0$. At the base of our domain, we impose magnetic stresses to vanish, setting $\partial B_x/\partial z = 0$ while at the top we consider a magnetic vacuum boundary condition, setting $B_x = 0$. We assume a constant axisymmetric heat flux $F_0$ coming from the hot planetary core, so that the bottom boundary condition for the axisymmetric part of the temperature $T_0(t)$ is ${\partial T_0}/{\partial z}={-F_0}/\bar\chi$. We use the flux $\sigma T_{\rm int}^4$, for an interior temperature of $T_{\rm int} = 150\ \mathrm{K}$, as in \citetalias{Menou2012b} and \citetalias{Hardy2023} for $F_0$. The axisymmetric interior flux thus impose $T_1$ and $T_2$ components to have constant zero flux, setting $\partial T_1 / \partial z = \partial T_2 / \partial z = 0$. At the outermost layer, we consider the thermal flux to adjust to the temperature. To do so we consider the thermal flux as
        \begin{equation}\label{eq:BC_flux_out}
            -\chi \frac{\partial T}{\partial z} = \frac{4}{3} \sigma T^4,
        \end{equation}
        where the $4/3$ factor is introduced to reproduce the results presented by \citet{Guillot2010}. With this equation, we can isolate the temperature derivative, and use the longitudinal expansion procedure to get $\partial T_0/\partial z$, $\partial T_1/\partial z$ and $\partial T_2/\partial z$. The free parameters used for the temperature profile are the equilibrium temperature $T_{\rm eq}$ and the thermal opacity $\kappa_{th}$.

        \begin{table}
            \begin{center}
                \begin{tabular}{ll}
                \hline\hline \\[-1em]
                Parameters & Values \\[0.2em]
                \hline \\[-1em]
                $B_0$ (G)                 & [1, 100] \\
                $\dot{v}$ (m s$^{-2}$)    & [0.0001, 0.01]   \\
                $\kappa_{\rm th}$ (m$^2$ kg$^{-1}$) & [0.0001, 0.001]  \\
                $T_{\rm eq}$ (K)          & [1000, 1200]    \\
                Pr                        & [0.01]    \\
                \hline
                \end{tabular}
            \end{center}
            \caption{\label{tab:parameters} Ranges of the parameters used in the simulations. See \citetalias{Hardy2023} for further justifications for the parameter ranges.}
        \end{table}
         
        We ran a grid of models covering the ranges of parameter values shown in Table \ref{tab:parameters}, using the same initial conditions for all simulations. For our initial conditions, we set the atmosphere to be in solid-body rotation, thus $u_x = 0$ across all layers, and the background magnetic field is undisturbed, leaving only the radial field, thus $B_x = 0$. We let the initial steady-state temperature profile obtained by an ODE solver relax in a time-dependent code, while also letting $P, \bar{\rho}, \bar{\mu}$ and $\bar{\chi}$ adjust to the changing temperature, but these quantities are kept constant once the relaxation procedure completed. As the fluid is initially under solid rotation, $T_2=0$ across the atmosphere. We then use Equation (\ref{eq:diffusivity}) with the initial $\tilde{T}(\phi)$ profile to construct the initial $\tilde{\eta}(\phi)$.

        \begin{figure}
            \centering
                \includegraphics[width=0.99\linewidth]{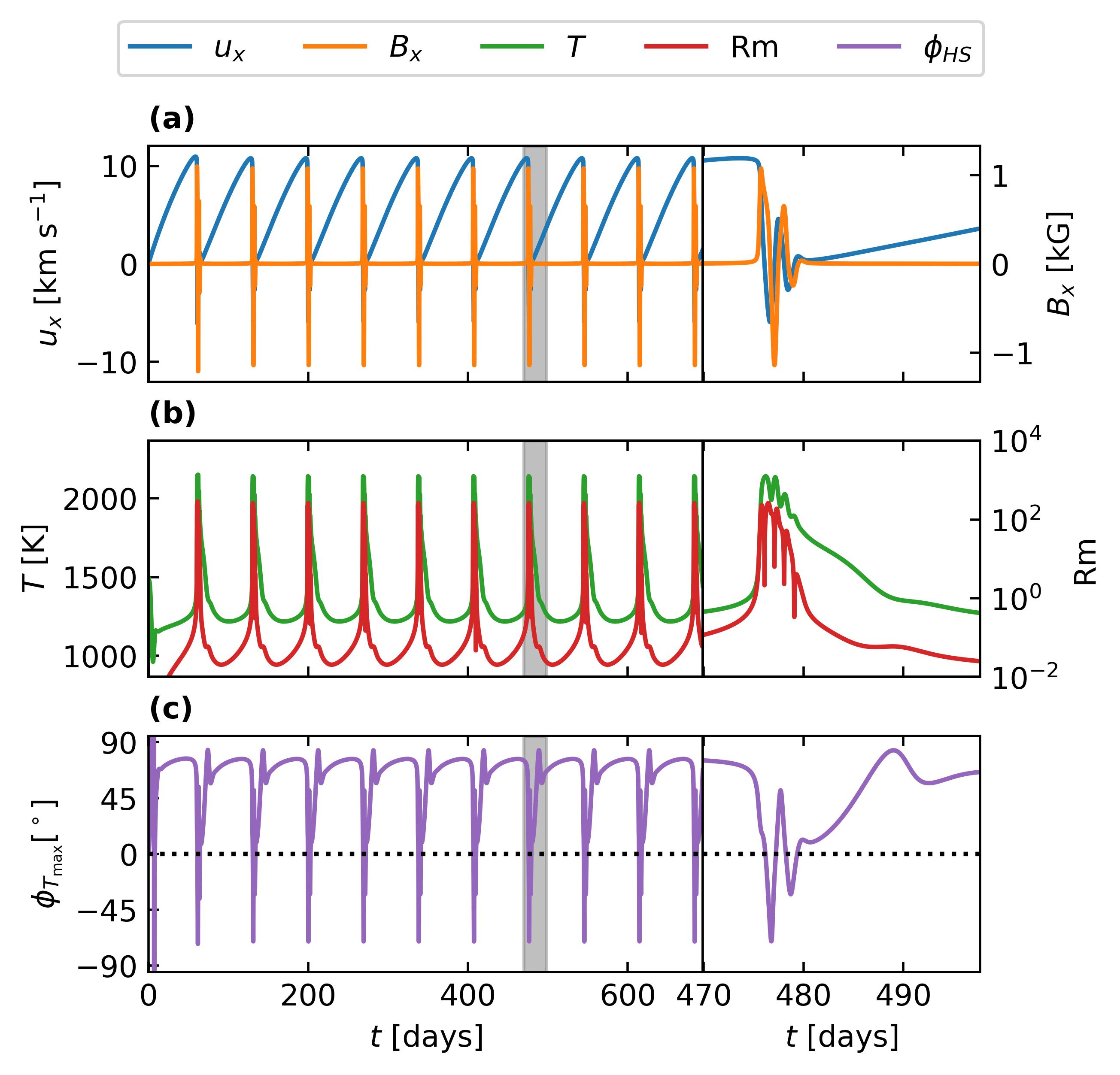}
                \caption{An example of bursting behavior driven by TRI. We show the time series of (a) velocity and magnetic field, (b) temperature and magnetic Reynolds number, and (c) angle of the maximum temperature. All quantities are measured in the center of the domain. The model shown has parameters $B_0=30$~G, $\dot{v}=0.006 \rm ~m~s^{-1}$, $\Pr=0.01$, $\kappa_{\rm th}=0.0008 \rm ~m^2~kg^{-1}$ and $T_{\rm eq}=1000$~K. The right panels zoom in on the gray shaded area indicated in the left panels.}
            \label{fig:time_series}
        \end{figure}

        \begin{figure*}[htpb]
            \centering
                \includegraphics[width=0.90\linewidth]{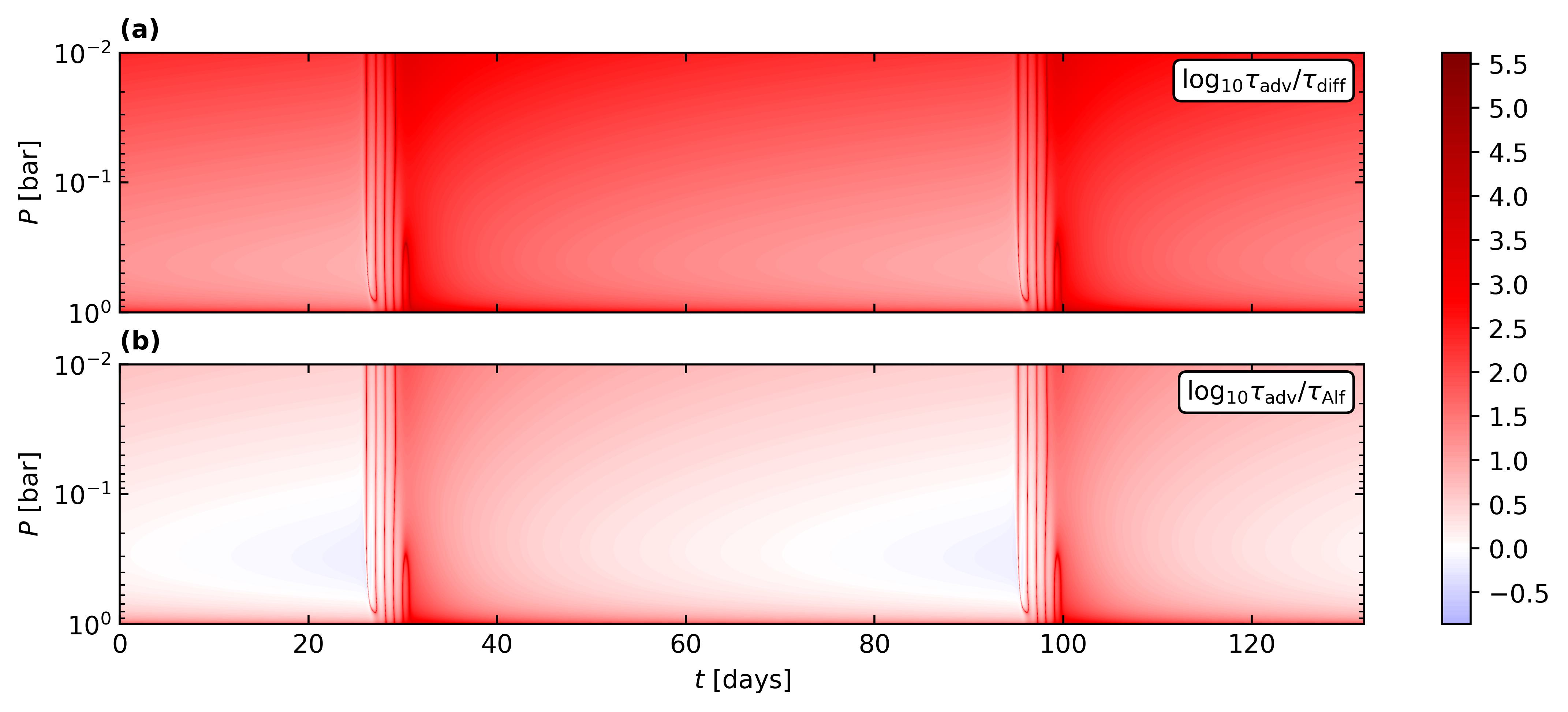}
                \caption{Ratios of timescales in the system during oscillations; (a) the advection timescale ($\tau_{\rm adv} = R/\abs{u_x}$) versus the thermal diffusion timescale ($\tau_{\rm diff} = H^2/\chi$), (b) the advection timescale versus the Alfv\'en timescale ($\tau_{\rm Alf}=H/u_A$). The length scale used to measure these is the pressure scale height, $H$, for $\tau_{\rm Alf}$ and $\tau_{\rm diff}$ which we update with the axisymmetric temperature value at every time step and the radius is used for $\tau_{\rm adv}$.}
            \label{fig:timescales}
        \end{figure*}

    \subsection{Numerical Scheme}\label{sec:numerical_scheme}

        The numerical solver is a straightforward generalization of the scheme presented in \citetalias{Hardy2023}. The velocity, the magnetic field, and all components of the temperature are advanced in time exactly as in \citetalias{Hardy2023}. We use 400 grid points equidistant in radius, and 360 grid points in $\phi$ to carry out the fit to $\eta(\phi)$. This numerical grid is fine enough to resolve boundary layers and longitudinal dependencies with good computational speed. We adopt a constant time step, chosen short enough to properly capture the bursts following the TRI. This requires $\Delta t = 20\ \mathrm{s}$, with a typical simulation requiring three million time steps.
        The radial resolution needs to be chosen carefully, because under-resolving the system leads to the appearance of kinks in the profiles that artificially inject energy into the system through enhanced dissipation. These kinks appear once the instability is triggered, therefore they do not fundamentally change the behavior of the system, but they do artificially extend the decay phase of post-burst Alfv\'en waves. 
        To avoid any numerical issue with our magnetic diffusivity, we have also imposed a maximum of $10^{12} \rm m^2~s^{-1}$ onto the MD, mostly notable near the beginning of the simulations, where the nightside is at its coldest.

        \begin{figure*}[htpb]
            \centering
                \includegraphics[width=0.85\linewidth]{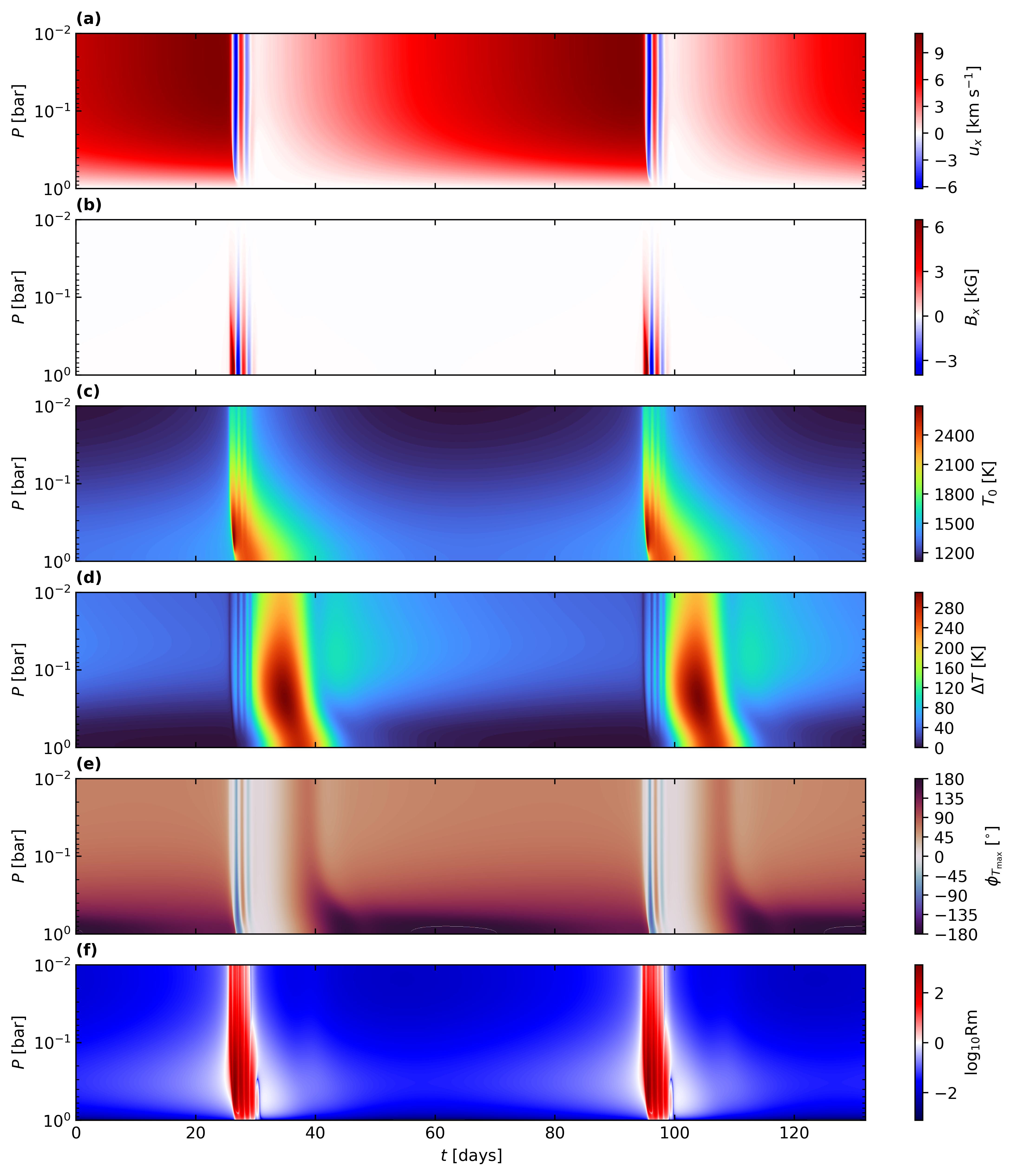}
                \caption{Spatiotemporal evolution of (a) the velocity, (b) the magnetic field, (c) temperature, (d) the longitudinal temperature difference between the hottest and coldest points at each depth, (e) angular offset of the hottest point at each depth, and (f) magnetic Reynolds number for the same simulation as in Figure \ref{fig:time_series}. 
                }
            \label{fig:butterfly}
        \end{figure*}
        
\section{Results}\label{sec:results}
    \subsection{A Representative Simulation}\label{sec:representative_sol}
     
        We first show a representative simulation exhibiting periodic bursts of Alfv\'en oscillations triggered by the TRI. To best illustrate the features of a burst and subsequent decaying Alfv\'en waves, we have chosen the simulation with parameters $B_0=30$~G, $\dot{v}=0.006 \rm ~m~s^{-1}$, $\Pr=0.01$, $\kappa_{\rm th}=0.0008 \rm ~m^2~kg^{-1}$ and $T_{\rm eq}=1000$~K. 
        
        Figure \ref{fig:time_series} shows the time series of the longitudinal velocity and magnetic field component, as well as the temperature and magnetic Reynolds number ($\Rm = u_x H / \eta_0$, with $H$ the local pressure scale height) at the center of the domain. In addition, we also show the longitudinal position of the maximum of $T(\phi)$ in the center of the domain (bottom panel). On the right-hand side of the figure, we zoom in to show the Alfv\'enic behavior of the velocity and magnetic field. The general behavior is similar to that found in  \citetalias{Hardy2023}. In panel (c) of Figure \ref{fig:time_series}, we see that the hottest point at mid-depth reaches longitudinal displacement of $76^{\circ}$ during the build-up phase. However, once the TRI is triggered, the position of the hottest point varies greatly, even reaching westward positions for over a day at a time.

        Figure \ref{fig:timescales} shows the ratios of the main timescales in the system during the simulation. Panel (a) indicates that thermal diffusion overwhelms advection almost throughout the simulation, with the exception of the Alfv\'enic oscillations at depth, where they share similar values. As thermal diffusion is almost always the leading heat transport mechanism to a varying degree throughout the simulation, finding the displacement of the temperature maximum is not simply a matter of time-integrating the velocity. As diffusion overwhelms advection, the hottest point does not reach large longitudes as heat diffuses quickly before it can be advected. In panel (b), comparing the advection and Alfv\'en timescales, we do not have a dominant timescale throughout. During build-up, advection dominates, but during and shortly after the TRI is triggered, the Alfv\'en timescale decreases to become slightly smaller than the advection time. The ratio of the Alfv\'en timescale to thermal diffusion timescale is not shown, as it is always $\gg 1$.

        \begin{figure}[htpb]
            \centering
                \includegraphics[width=1.0\linewidth]{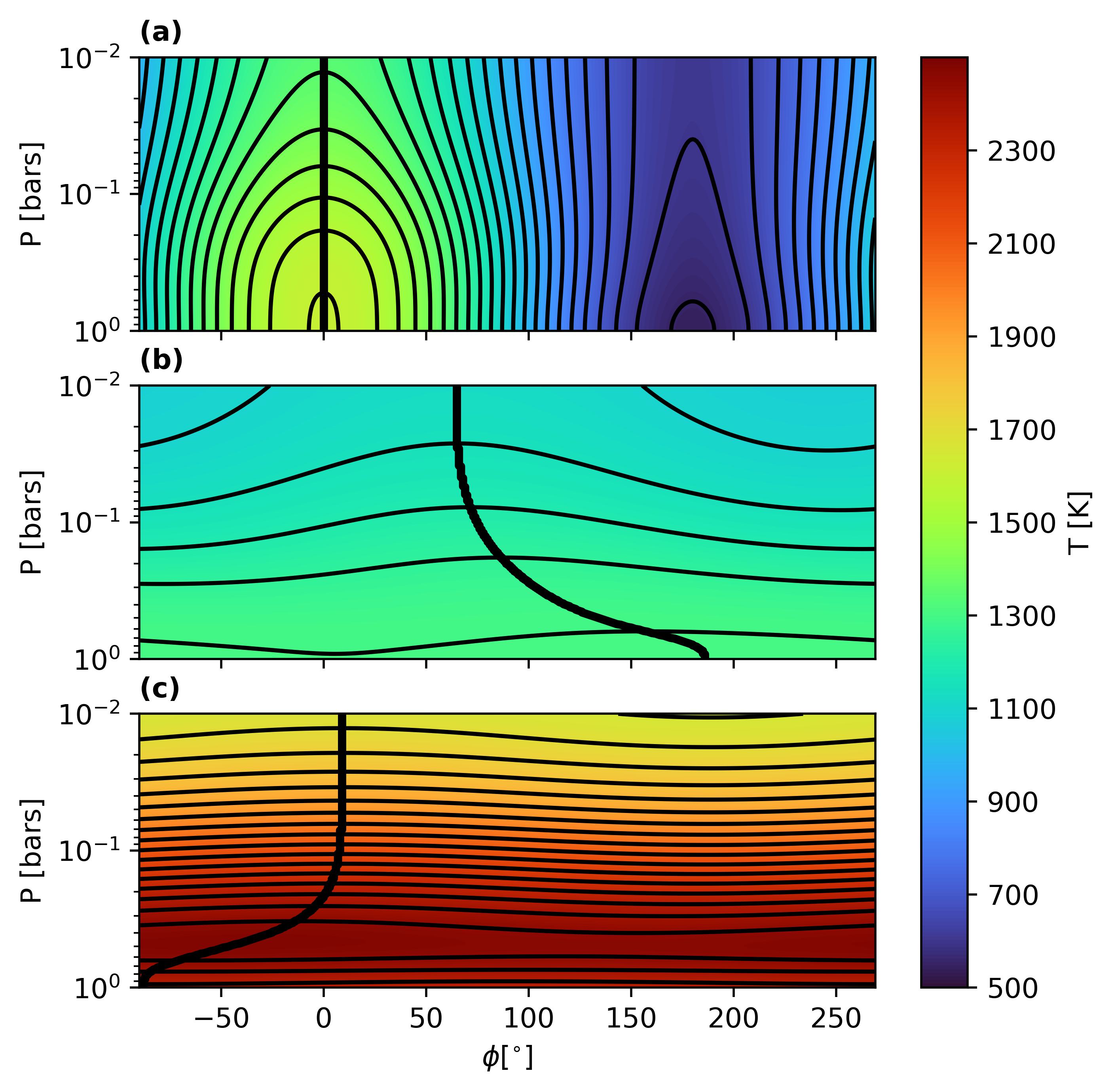}
                \caption{Depth-longitude temperature maps during three different instants; (a) initial condition, (b) first build-up phase and (c) during the first burst of the simulation. The thin black lines each represent an isocontour separated by 50~K from each others, and the thicker black line track the hottest point at each layers.}
            \label{fig:temperature_map}
        \end{figure}
        
        Figure \ref{fig:butterfly} shows the depth-dependence of key variables as a function of time. Panels (a), (b) and (c) show the evolution of the velocity, magnetic field and the temperature. These are similar to the results from \citetalias{Hardy2023}. We see strong heating of the atmosphere at depth during the oscillatory phase, with the TRI initially starting at a pressure of $\approx 0.2$ bars and propagating downwards towards the 1 bar level. Panel (f) shows the evolution of the magnetic Reynolds number. As the magnetic Reynolds number crosses unity, the TRI is triggered and as it dissipates through Alfv\'enic oscillations, the system returns to a quiescent build-up phase evolving towards a steady state, until the next TRI is triggered. The magnetic field strength during the oscillation phase can become large enough that magnetic pressure is significant in some regions; we discuss this further in Appendix B.

        The longitudinal variations of temperature are summarized in panel (d), which shows the temperature difference between coldest and hottest spots at each depth, and panel (e), which shows the angular position $\phi_{T_\mathrm{max}}$ of the hottest point at each depth. The rapid vertical thermal diffusion prevents strong variations of $\phi_{T_\mathrm{max}}$ with depth, except at the highest pressures where the angular displacements lag those at higher altitudes.
        The largest offsets occur when the advection timescale is closest in size to the thermal diffusion timescale (Figure \ref{fig:timescales}). Were advection to be the dominant heat transport mechanism, we would expect an isothermal ring to form around the planet; in the opposite limit of negligible advection, we would expect the hottest point to remain at the substellar point. As the simulation moves between these limits, the offset can go beyond the terminator when advection is strongest, but returns on the eastern dayside when diffusion dominates again. The largest temperature difference across the surface is seen in the decaying phase, where $\Delta T$ reaches $\approx 310\ \mathrm{K}$ at $P\approx 0.2$ bars, lasting for about 10 days after the burst of oscillations. This is much smaller than the change in the axisymmetric component of the temperature $T_0$ during the burst, so that the heating from the TRI is close to being axisymmetric.

        Figure \ref{fig:temperature_map} shows the depth-longitude temperature maps during three distinct instants. Panel (a) shows the initial condition of the atmosphere. We see a temperature contrast of about 1000~K as the fluid is at rest. In panel (b), the advection has become important and the temperature is almost homogeneous in longitude. The hot spot is also advected eastward, with the smallest deviation at lower pressure, where the thermal diffusion is fastest, as shown in Figure \ref{fig:timescales}. Panel (c) shows the large rise in temperature caused by the TRI and the shift in hot spot cause by the reversal of the wind. While at lower pressure, the hot spot is near substellar, while at depth the hottest region is well into the western side of the dayside of the planet. 

\begin{figure}
            \centering
                \includegraphics[width=0.99\linewidth]{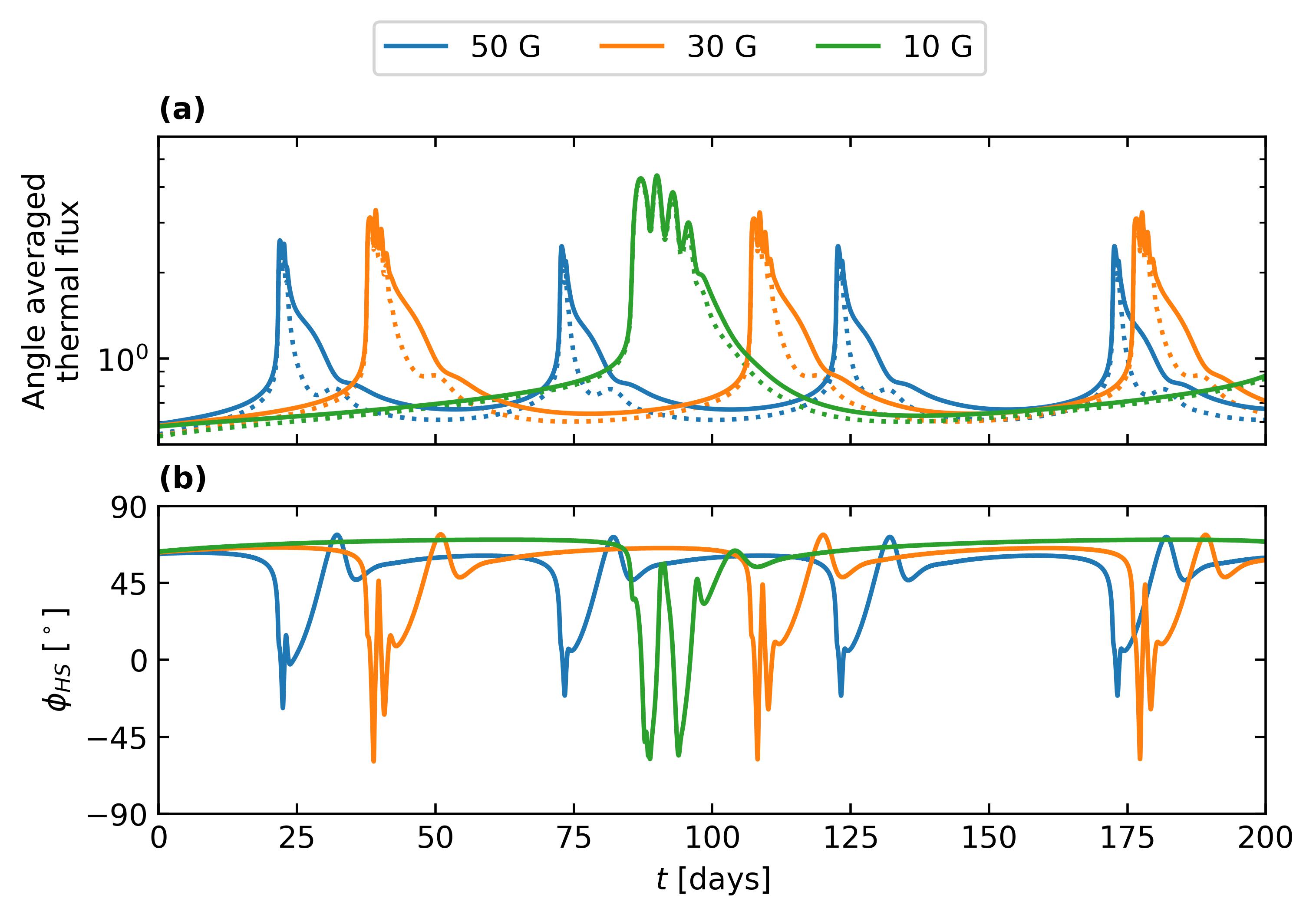}
                \caption{Day and nightside angle-averaged thermal flux  at the surface (full and dotted lines respectively), normalized by the flux value on the dayside at the start of the simulation (top panel), and the angular offset of the maximum in the thermal flux (bottom panel). We show results for simulations with a range of different $B_0$ values. The representative solution from \S \ref{sec:representative_sol} ($B_0=30\ \mathrm{G}$) is plotted in orange.}
            \label{fig:heat_flux}
        \end{figure}

        Figure \ref{fig:heat_flux} (a) shows the time-evolution of the thermal flux coming out of the atmosphere for different values of $B_0$. We first evaluate the outwards flux at the surface $F_\mathrm{out}(\phi)=-\bar{\chi}\partial \tilde{T}(\phi)/\partial z|_{z=z_\mathrm{top}}$ as a function of $\phi$, and then integrate $\int_{-\pi/2}^{\pi/2} F_\mathrm{out}(\phi) \cos\phi\ d\phi$ to approximate observing the dayside of the planet (e.g.~during secondary eclipse). The same procedure is used for the nightside.
        The kinetic energy from the forcing, which is ultimately transformed into heat through magnetic field induction followed by ohmic heating, enhances the thermal flux by factors of $2$--$3$ compared to the initial advection free value. The dotted lines representing the angle-averaged thermal flux of the nightside is indistinguishable from the dayside at the start of TRI, but as irradiation is smaller, this region can cool down faster. 
        
        Figure \ref{fig:heat_flux} (b) shows the longitudinal position of the thermal flux peak. The faster vertical thermal diffusion at low pressure means that the offset of the hot spot at the surface is reduced compared to the offset of the hottest point at depth (Figures \ref{fig:butterfly} and \ref{fig:temperature_map}). Nevertheless, the behavior in time remains similar: it grows toward an equilibrium displacement of $60$--$70^\circ$ eastwards during the quiescent phase, oscillates around $0^\circ$ after the onset of TRI, reaching in some cases more than $50^\circ$ westward. Once the flow decouples from the field and the oscillations damp down, the hot spot displacement then relaxes back toward equilibrium over about 10 days.
        
    \subsection{Dependence on input parameters}\label{sec:represent}       

        The set of parameters chosen for the simulation just discussed are only one of many choices that lead to TRI. With four different control parameters that can be varied, the behavior of the different solutions changes. Increasing the radial magnetic field strength reduces the recurrence period of the TRI, by reducing the velocity needed to trigger the instability, heating being faster with a stronger magnetic field. As such, the kinetic energy reservoir is smaller when the TRI triggers for larger $B_0$ values, thus the thermal flux outgoing from the atmosphere is reduced, as seen in Figure \ref{fig:heat_flux} (a). The hot spot offset gets smaller with stronger $B_0$ values, as seen in Figure \ref{fig:heat_flux} (b), reflecting the quenching action of the Lorentz force on the zonal flow. However, if the field is too strong, then the system simply reaches steady-state, as shown in \citetalias{Hardy2023}. Increasing $B_0$ also reduces the number of Alfv\'en oscillations as the field is stiffer. 
        The acceleration $\dot{v}$, when increased, also reduces the recurrence time as more energy is injected into the system per unit time. This energy is converted into heat and the critical temperature where $\Rm = 1$ is reached faster. However, if the acceleration is too large, then the system will remain too hot and recurrent TRI will not be possible; whereas if it is too small, then the critical temperature will never be reached. The equilibrium temperature dictates the initial magnetic Reynolds number value. Considering only equilibrium values compatible with TRI, a hotter atmosphere will reach TRI faster than a colder one with the same parameters, therefore reducing the recurrence period. 
        As for the thermal opacity $\kappa_{\rm th}$, a larger value will also help the atmosphere to reach TRI faster as heat is dissipated more slowly, therefore also reducing the recurrence period. The quantitative effect of $B_0$ and $T_\mathrm{eq}$ can be seen in Figure 10 of \citetalias{Hardy2023}. 

        Although we find very similar behavior to \citetalias{Hardy2023}, for a given set of parameters the precise evolution is different. As discussed in section \ref{sec:model}, we use the longitudinally-averaged MD $\eta_0$ in the induction equation, which is larger than in the fully axisymmetric model of \citetalias{Hardy2023}, where only substellar conditions were considered. As we are also considering the colder nightside, $\eta_0$ is larger than $\eta$ at the substellar point. Thus, the system needs to reach larger velocities to generate large enough values of $B_x$ to compensate for the larger diffusivity, which reduces the value of the magnetic Reynolds number, making it harder for the system to reach the critical value of $\Rm = 1$, triggering the instability. Only if the system is able to reduce its temperature after the TRI through an outgoing thermal flux at the upper boundary, thus going back to Rm smaller than unity, can the system generate recurrent bursts. Overall, the larger MD shifts the instability region to larger values of $T_{\rm eq}$, $\kappa_{\rm th}$ and $\dot{v}$.

\section{Discussion}\label{sec:discussion}

\begin{figure*}
        \centering
            \includegraphics[width=0.80\linewidth]{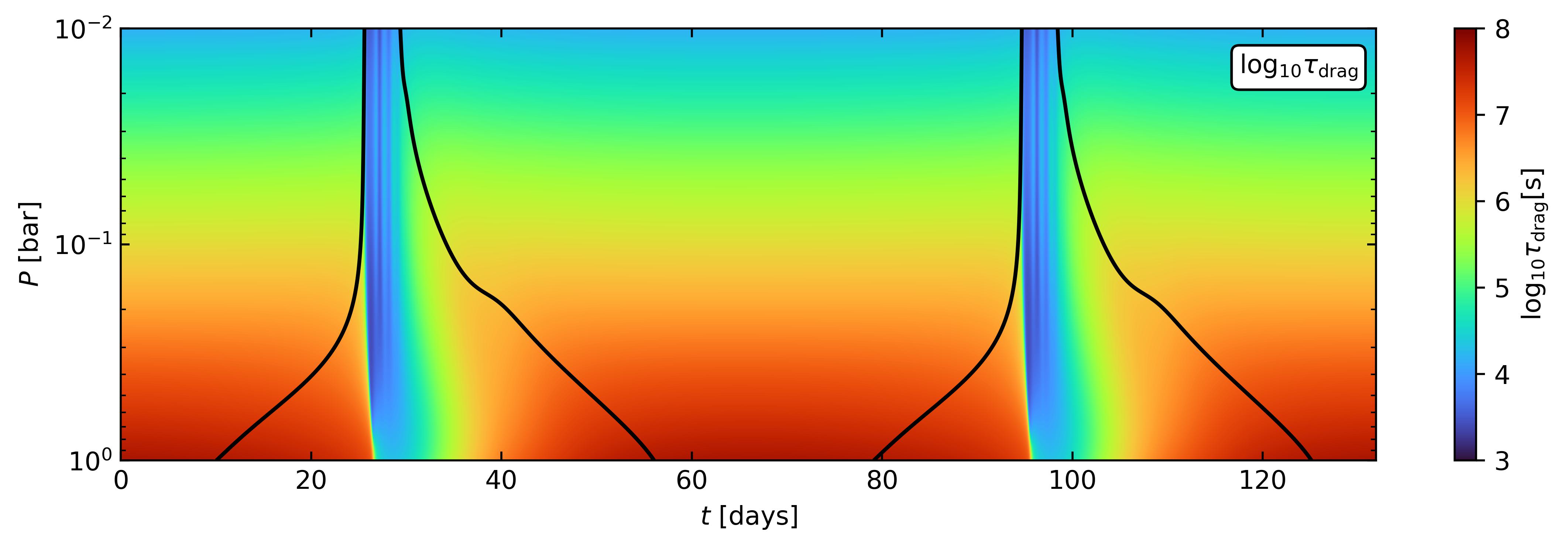}
            \caption{Total drag timescale defined as $\tau_{\rm drag} = [\tau_\nu^{-1} + \tau_{\rm mag}^{-1}]^{-1}$, where $\tau_\nu=H^2/\nu$ is the viscous drag timescale and $\tau_{\rm mag} = \mu_0\rho\eta_0/B_0^2$ is the magnetic drag timescale. Viscous drag timescale ($\tau_\nu$) is shorter during build-up, but as TRI occurs, $\tau_{\rm mag}$ becomes much smaller and the magnetic drag timescale dominates the flow dynamics. The black line denotes where $\tau_\nu=\tau_{\rm mag}$.}
        \label{fig:drag_timescale}
    \end{figure*}

    We have presented here a simple model which encapsulates the longitudinal dependency of the temperature in a HJ atmosphere susceptible to the TRI. Building on the works of \citetalias{Hardy2023}, this model has allowed us to characterize the reaction of the hot spot offset to our imposed acceleration and the TRI. We were able to simulate the hot spot displacement during a burst triggered by TRI from 0.01 bar to 1 bar. These simulations suggest that it may be possible to detect the occurrence of this instability in the atmospheres of these gas giants. Our simulations indicate that the TRI could produce significant hot spot offset variations, although somewhat larger than the offset range found from \textit{Spitzer}'s phase curves by \citet{Bell2021}. Indeed, the simulation presented in this work showed peak flux offset ranging from $66^{\circ}$ eastward to $56^{\circ}$ westward. The recurrence time and Alfv\'enic oscillation periods found in \citetalias{Hardy2023} are still valid for this longitudinally-extended model, with our representative solution being on the shorter end of the period spectrum. The reanalysis of phase curves of 16 planets done by \citet{Bell2021} shows offsets between $-38.7^{\circ+3.2}_{~-3.2}$ (CoRoT-2b) and $43.4^{\circ+5.4}_{~-6.1}$ (HD209458b) in the colder regime ($T_{\rm irr}< 2500$~K). Comparing our results to the model presented in \citet{Komacek2017}, our offset results for oscillating systems are comparable to the fastest superrotating equatorial jets presented in that study ($8 \rm  ~km~s^{-1}$), and long drag timescales ($\tau_{\rm drag} = 10^6-10^7$~s). Indeed, \citet{Komacek2017} are expecting offsets of around $60-70^{\circ}$ with these physical characteristics. Figure \ref{fig:drag_timescale} shows drag timescale throughout the representative solution. Outside of the TRI, the viscous drag is dominating, with associated timescale in the range $10^5-10^8$~s, depending on depth. During TRI, magnetic drag is the biggest contributor, dropping the drag timescale down to $10^3-10^4$~s. The repercussions of this drop is seen in the hot spot offset. Indeed, after the decaying oscillations and before build-up starts, the offset is near $0^{\circ}$, as predicted by models with small drag timescales in \citet{Komacek2017}. Thus, our offset values seem reasonable considering our simplified model.

    \citet{Keating2019} have inferred the temperatures on the day and nightside for 12 exoplanets. The temperature contrast between the dayside and the nightside varies between 150~K for HD189733b and 1040~K for WASP-18b. We note however that only HD189733b and HD209458b have dayside temperatures below 1500~K. The latter has a temperature contrast of 189~K. The rest of the sample presented in that paper is too hot to be conducive to the TRI. We note that the analysis presented by \citet{Bell2021} find a temperature contrast of 287~K for HD189733b. The representative solution of Section \ref{sec:representative_sol} has a maximum temperature contrast of 310~K after the TRI burst. However, during build-up as the system nears steady-state before the TRI triggers, the temperature contrast is around 50~K only. Thus, assuming the temperature contrast of \citet{Keating2019} are from steady-state planets, our model seems to underestimate this characteristic of the atmosphere. Our first order Fourier expansion restricts the way we can heat the atmosphere in our modeling. As our heating is modulated by $(1+\sin(\phi))/2$, we are also heating parts of the nightside, reducing the temperature contrast between day and nightside.

    The flux variation caused by the thermal runaway of the TRI, as shown in Figure \ref{fig:heat_flux}, would certainly be the most conspicuous observable telltale sign of the TRI. Indeed, for our representative solution, the outward thermal flux momentarily grows fivefold. Therefore, any planets with a much larger luminosity than expected may be undergoing TRI. However, the short-comings of our model must again be emphasized. The dynamical variables, $u_x$ and $B_x$, are assumed axisymmetric. The longitudinal velocity is expected to have many small scale structures interacting with higher and lower latitudes, breaking symmetry. Large scale flows coming from higher and lower latitudes would also break symmetry. As for the magnetic field, only a perfectly aligned dipole would be axisymmetric. With such asymmetry, we would assume to be harder for the TRI to trigger globally as turbulence, local instabilities, etc. may slow down winds, thus reducing ohmic heating. On the other hand, turbulence may enhance ohmic heating through shearing. Thus, the exact effects of a more complex model on the TRI is unclear. TRI may rather operate locally and significantly reduce the predicted values of Figure \ref{fig:heat_flux}. In addition, while the model presented in this work requires ohmic heating to reach the critical magnetic Reynolds number, such heating could be secondary if considering the increase in temperature at the morning terminator. Indeed, as the cold gas of the nightside arrives on the dayside, stellar flux could be the heat source required to trigger TRI at the morning terminator.

    Simulations with larger magnetic field strengths tend to show shorter recurrence periods and smaller oscillation amplitudes, and vice versa, for a given set of parameters. From \citetalias{Hardy2023}, we recall that beyond a certain magnetic field strength, for a given set of $T_{\rm eq}$, $\dot{v}$ and $\kappa_{\rm th}$, the system is no longer susceptible to the TRI, and instead reaches a steady state. Therefore, as the field strength increases, the offset range diminishes. Moreover, as argued in Section \ref{sec:represent}, the other observable characteristic of TRI, the outgoing flux, should also favor weaker field, as stronger fields show smaller variations in their outgoing thermal flux. The estimated dipole field strength by \citet{Yadav2017} for HJs averages around 100~G, but as our input field strength represent the radial component, and as argued in \citetalias{Hardy2023}, a 100~G dipole field yields a radial component up to 17.4~G with an inclination of $10^{\circ}$ in respect to the rotation axis. At the maximum value predicted by \citet{Yadav2017}, $\approx250$~G, and the same inclination, we would get a radial field strength of 43.4~G. Thus, given the right set of parameters, all field strength proposed by \citet{Yadav2017} are potential candidates for TRI.

    As the true physical boundary conditions on the different temperature component is unclear, we have tested many different combinations, before settling on the one presented in this work. We tried having $T_1$ and $T_2$ fixed to zero at both boundaries. At the bottom, this would represent a core with an axisymmetric temperature. At the top, it would mimic an efficient thermal conductor. For the axisymmetric part, we kept it constant at the top and kept the internal planetary flux, as in \citetalias{Hardy2023}. However, with these assumptions, the temperature contrast had a maximum around half that characterizing our representative simulation, i.e., much smaller than anticipated from other numerical models and observations. Keeping the temperature constant at the outermost layer also has the effect of radically increasing the outgoing thermal flux. The angle averaged thermal flux coming out of the atmosphere would increase tenfold during TRI. Thus, the flux-temperature relation was chosen for the top boundary condition on the temperature components, and an insulating bottom boundary condition for $T_1$ and $T_2$ was also adopted.

    While we only presented results from simulations with Prandtl number of 0.01, we tested larger values, i.e. giving more importance to viscosity. A larger Prandtl number has the effect of reducing the drag timescale, thus reducing the flow velocity, but also enhancing viscous heating. This shifts the region of parameter space susceptible to TRI to lower temperature and stronger forcing, as to counteract the effect of stronger viscous drag. However, velocities of around $10~\rm km~s^{-1}$ seemed to still be required to trigger TRI. For example, Pr=0.1 yields periodic TRI at equilibrium temperatures of 700-800~K, putting these atmospheres closer to the warm Jupiters regime rather than the hot Jupiters one. Moreover, enhancing the Prandtl number beyond the expected value from micro physics could also be a way to take into account omitted physical phenomenon, such as flow out of the equatorial plane, shocks, and turbulence.

    Finally, we want to point out that the magnetic Prandtl number (${\rm Pm}=\nu/\eta_0$) always remain below unity. For the representative solution of Section \ref{sec:representative_sol}, it reaches a maximum value of $1.4\times 10^{-1}$, and goes down to $\approx 10^{-7}$ at depth during the quiescent phase, when $\eta_0$ is at its largest. Thus, ohmic dissipation is always greater than viscous dissipation.

    In summary, together with \citetalias{Hardy2022} and \citetalias{Hardy2023}, our results show that a temperature dependent MD can impact the atmospheric dynamics drastically. Thus, in the regimes where the magnetic Reynolds number is around unity, it would be imperative to further extend models to implement such dependence. A complete 3D model or simulation would be necessary to incorporate all physical mechanisms susceptible to impact, positively or negatively, the TRI. Such a model would also lead to better predictions of the observable offset variability caused by the TRI, but also the changes in luminosity associated with the thermal runaway of the instability.

\vspace{\baselineskip}
    This work was supported by the Natural Sciences and Engineering Research Council of Canada (NSERC) Discovery grants RGPIN-2018-05278, RGPIN-2023-03620, and RGPIN-2024-04050. R.H., P.C., and A.C. are members of the Centre de Recherche en Astrophysique du Qu\'ebec (CRAQ). A.C.~is grateful to the Isaac Newton Institute for Mathematical Sciences, Cambridge, for support and hospitality during the programme ``Anti-diffusive Dynamics: from sub-cellular to astrophysical scales'' supported by EPSRC grant no EP/R014604/1.

\appendix

\section{Accuracy of the sinusoidal assumption for the angular dependence of the magnetic diffusivity}\label{appendix:diffusivity}

    As discussed in the main text, because $\eta(T)$ is strongly non-linear in $T$ (Equation~(\ref{eq:diffusivity})), a sinusoidal $T(\phi)$ as assumed in Equation (\ref{eq:temperature_expansion}) does not give $\eta(\phi)$ that is also sinusoidal. Therefore, when we expand $\eta(\phi)$ in the form of Equation (\ref{eq:eta_expansion}), with $\sin\phi$ and $\cos\phi$ terms only, we introduce an inaccuracy. 
    
\begin{figure*}
        \centering
            \includegraphics[width=0.99\linewidth]{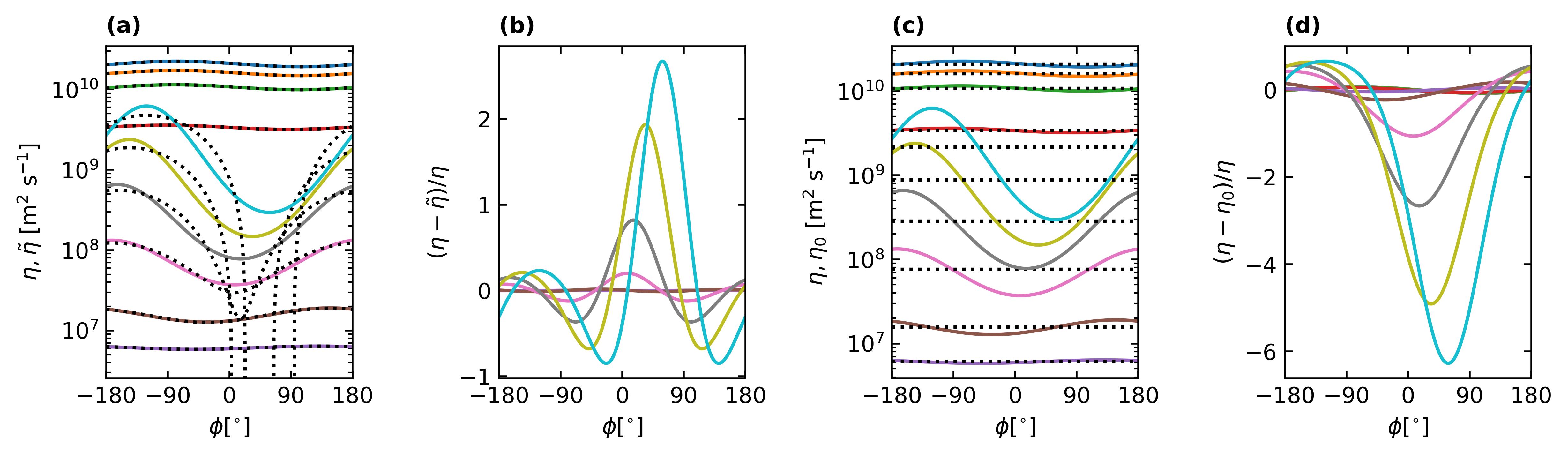}
            \caption{(a) $\eta(\phi)$ corresponding to $T(\phi)$ obtained using Equation (\ref{eq:diffusivity}) (solid lines in color) and the corresponding Fourier fit $\tilde{\eta}(\phi)$ (Eq.~\ref{eq:eta_expansion}) (black dotted lines).
            (b) The fractional difference between $\eta(\phi)$ and $\tilde{\eta}(\phi)$. The values are taken equidistant in time during a burst and the following decaying Alfv\'en waves (right-hand side of Figure \ref{fig:time_series}). (c) Similar to panel (a) but now comparing $\eta(\phi)$ with the longitudinal average $\eta_0$.
            (d) The fractional difference between $\eta(\phi)$ and $\eta_0$.}
        \label{fig:magnetic_diffusivity_error}
    \end{figure*}

    To assess the magnitude of this error, Panel (a) of Figure \ref{fig:magnetic_diffusivity_error} compares the true $\eta(\phi)$ corresponding to $T(\phi)$ obtained using Equation (\ref{eq:diffusivity}) (solid lines in color) and the corresponding Fourier fit $\tilde{\eta}(\phi)$ (Equation~(\ref{eq:eta_expansion})) (black dotted lines). We show curves at various times during the oscillation cycle of the TRI, during which $\eta$ varies by more than three orders of magnitude as the atmospheric temperature changes. We note how at some epochs, the fitted diffusivity may reach negative values at some longitudes. These negative values happen when the range of the diffusivity is largest, making it difficult to fit a sinusoidal curve onto it. The relative deviations between the real profiles and their fit are plotted in panel (b). Considering the large range over which $\eta$ varies over the oscillation, we find that the $\eta(\phi)$ profiles are reproduced reasonably by the sinusoidal approximation, with largest deviations reaching 250\% locally, but usually remaining around 20\%. 

    Panels (c) and (d) of Figure \ref{fig:magnetic_diffusivity_error} are similar to panels (a) and (b) except now we compare the longitudinally-averaged component $\eta_0$ with the true $\eta(\phi)$ profile. This is important because, as discussed in section \ref{sec:model}, we use $\eta_0$ to replace $\eta$ in the induction equation. We underestimate the true $\eta$ by more than a factor of six at the substellar point at some phases of the Alfv\'enic oscillations, down to 8\% right before TRI. This leads to a small change in the instability boundary in parameter space compared to \citetalias{Hardy2023}, as discussed in section \ref{sec:represent}.

\section{Ratio of magnetic pressure to gas pressure}\label{appendix:buoyancy}

\begin{figure*}
        \centering
            \includegraphics[width=0.99\linewidth]{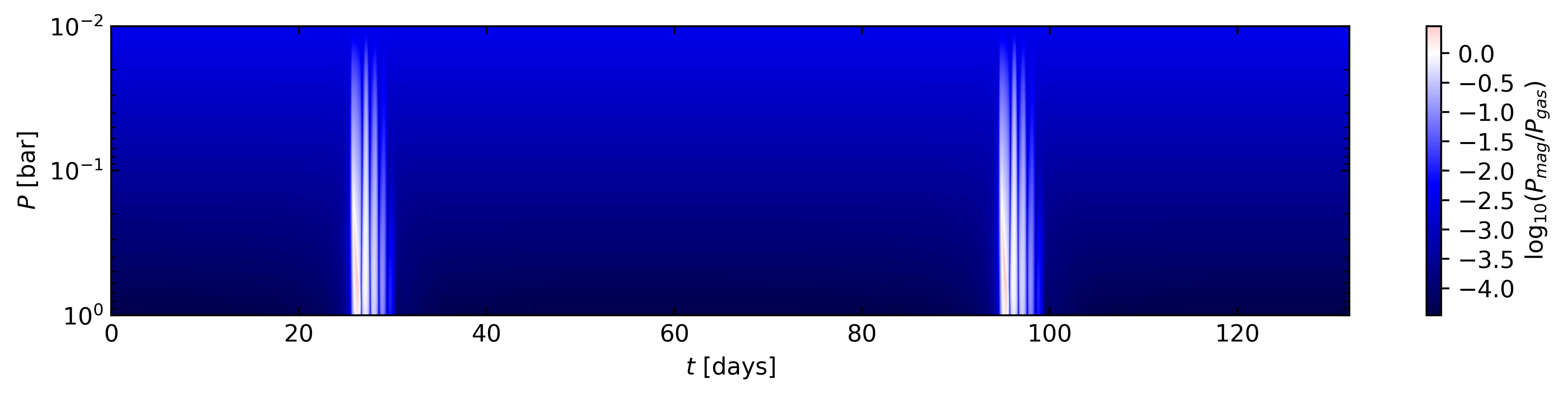}
            \caption{Ratio of magnetic to gas pressure in $\log_{10}$ as a function of time and pressure, for the solution presented in Section \ref{sec:representative_sol}.}
        \label{fig:magnetic_pressure}
    \end{figure*}

    Because our model precludes vertical motions, there is no consequence to the magnetic pressure $P_\mathrm{mag}$ exceeding the gas pressure $P_\mathrm{gas}$. In reality this could possibly lead to a buoyancy-driven instability (e.g.~\citealt{Newcomb1961,Parker1975,Fan2021}; see \citetalias{Hardy2023} for further discussion). We do find that many models reach $P_\mathrm{mag}/P_\mathrm{gas}$ ratios larger than unity in some small regions of the atmosphere.
    As an example, Figure \ref{fig:magnetic_pressure} shows $P_\mathrm{mag}/P_\mathrm{gas}$ (the inverse plasma-$\beta$) for the representative model discussed in Section \ref{sec:results}. It can be seen that $P_\mathrm{mag}/P_\mathrm{gas}$ reaches $\approx 3$ near the base of the layer during the Alfven oscillations. Investigation of the effects of this, for example whether instabilities have time to grow and the subsequent vertical transport, would be an interesting avenue for future investigation.

\bibliography{zmain}{}

\begin{thebibliography}{}
\expandafter\ifx\csname natexlab\endcsname\relax\def\natexlab#1{#1}\fi
\providecommand{\url}[1]{\href{#1}{#1}}
\providecommand{\dodoi}[1]{doi:~\href{http://doi.org/#1}{\nolinkurl{#1}}}
\providecommand{\doeprint}[1]{\href{http://ascl.net/#1}{\nolinkurl{http://ascl.net/#1}}}
\providecommand{\doarXiv}[1]{\href{https://arxiv.org/abs/#1}{\nolinkurl{https://arxiv.org/abs/#1}}}

\bibitem[{{Armstrong} {et~al.}(2016){Armstrong}, {de Mooij}, {Barstow},
  {Osborn}, {Blake}, \& {Saniee}}]{Armstrong2016}
{Armstrong}, D.~J., {de Mooij}, E., {Barstow}, J., {et~al.} 2016, Nature
  Astronomy, 1, 0004, \dodoi{10.1038/s41550-016-0004}

\bibitem[{{Batygin} \& {Stevenson}(2010)}]{Batygin2010}
{Batygin}, K., \& {Stevenson}, D.~J. 2010, \apjl, 714, L238,
  \dodoi{10.1088/2041-8205/714/2/L238}

\bibitem[{{Bell} {et~al.}(2019){Bell}, {Zhang}, {Cubillos}, {Dang}, {Fossati},
  {Todorov}, {Cowan}, {Deming}, {Zellem}, {Stevenson}, {Crossfield},
  {Dobbs-Dixon}, {Fortney}, {Knutson}, \& {Line}}]{Bell2019}
{Bell}, T.~J., {Zhang}, M., {Cubillos}, P.~E., {et~al.} 2019, \mnras, 489,
  1995, \dodoi{10.1093/mnras/stz2018}

\bibitem[{{Bell} {et~al.}(2021){Bell}, {Dang}, {Cowan}, {Bean}, {D{\'e}sert},
  {Fortney}, {Keating}, {Kempton}, {Kreidberg}, {Line}, {Mansfield},
  {Parmentier}, {Stevenson}, {Swain}, \& {Zellem}}]{Bell2021}
{Bell}, T.~J., {Dang}, L., {Cowan}, N.~B., {et~al.} 2021, \mnras, 504, 3316,
  \dodoi{10.1093/mnras/stab1027}

\bibitem[{{Cooper} \& {Showman}(2005)}]{Cooper2005}
{Cooper}, C.~S., \& {Showman}, A.~P. 2005, \apjl, 629, L45,
  \dodoi{10.1086/444354}

\bibitem[{{Dang} {et~al.}(2018){Dang}, {Cowan}, {Schwartz}, {Rauscher},
  {Zhang}, {Knutson}, {Line}, {Dobbs-Dixon}, {Deming}, {Sundararajan},
  {Fortney}, \& {Zhao}}]{Dang2018}
{Dang}, L., {Cowan}, N.~B., {Schwartz}, J.~C., {et~al.} 2018, Nature Astronomy,
  2, 220, \dodoi{10.1038/s41550-017-0351-6}

\bibitem[{{Davidson}(2001)}]{Davidson2001}
{Davidson}, P.~A. 2001, {An Introduction to Magnetohydrodynamics} ({Cambridge
  University Press})

\bibitem[{{Dietrich} {et~al.}(2022){Dietrich}, {Kumar}, {Poser}, {French},
  {Nettelmann}, {Redmer}, \& {Wicht}}]{Dietrich2022}
{Dietrich}, W., {Kumar}, S., {Poser}, A.~J., {et~al.} 2022, \mnras, 517, 3113,
  \dodoi{10.1093/mnras/stac2849}

\bibitem[{{Draine} {et~al.}(1983){Draine}, {Roberge}, \&
  {Dalgarno}}]{Draine1983}
{Draine}, B.~T., {Roberge}, W.~G., \& {Dalgarno}, A. 1983, \apj, 264, 485,
  \dodoi{10.1086/160617}

\bibitem[{{Fan}(2021)}]{Fan2021}
{Fan}, Y. 2021, Living Reviews in Solar Physics, 18, 5,
  \dodoi{10.1007/s41116-021-00031-2}

\bibitem[{{Guillot}(2010)}]{Guillot2010}
{Guillot}, T. 2010, \aap, 520, A27, \dodoi{10.1051/0004-6361/200913396}

\bibitem[{{Hardy} {et~al.}(2023){Hardy}, {Charbonneau}, \&
  {Cumming}}]{Hardy2023}
{Hardy}, R., {Charbonneau}, P., \& {Cumming}, A. 2023, \apj, 959, 41,
  \dodoi{10.3847/1538-4357/ad0968}

\bibitem[{{Hardy} {et~al.}(2022){Hardy}, {Cumming}, \&
  {Charbonneau}}]{Hardy2022}
{Hardy}, R., {Cumming}, A., \& {Charbonneau}, P. 2022, \apj, 940, 123,
  \dodoi{10.3847/1538-4357/ac9bfc}

\bibitem[{{Hindle} {et~al.}(2019){Hindle}, {Bushby}, \& {Rogers}}]{Hindle2019}
{Hindle}, A.~W., {Bushby}, P.~J., \& {Rogers}, T.~M. 2019, \apjl, 872, L27,
  \dodoi{10.3847/2041-8213/ab05dd}

\bibitem[{{Hindle} {et~al.}(2021{\natexlab{a}}){Hindle}, {Bushby}, \&
  {Rogers}}]{Hindle2021b}
---. 2021{\natexlab{a}}, \apj, 922, 176, \dodoi{10.3847/1538-4357/ac0e2e}

\bibitem[{{Hindle} {et~al.}(2021{\natexlab{b}}){Hindle}, {Bushby}, \&
  {Rogers}}]{Hindle2021a}
---. 2021{\natexlab{b}}, \apjl, 916, L8, \dodoi{10.3847/2041-8213/ac0fec}

\bibitem[{{Hubbard} {et~al.}(2012){Hubbard}, {McNally}, \& {Mac
  Low}}]{Hubbard2012}
{Hubbard}, A., {McNally}, C.~P., \& {Mac Low}, M.-M. 2012, \apj, 761, 58,
  \dodoi{10.1088/0004-637X/761/1/58}

\bibitem[{{Imamura} {et~al.}(2020){Imamura}, {Mitchell}, {Lebonnois}, {Kaspi},
  {Showman}, \& {Korablev}}]{Imamura2020}
{Imamura}, T., {Mitchell}, J., {Lebonnois}, S., {et~al.} 2020, \ssr, 216, 87,
  \dodoi{10.1007/s11214-020-00703-9}

\bibitem[{{Jackson} {et~al.}(2019){Jackson}, {Adams}, {Sandidge}, {Kreyche}, \&
  {Briggs}}]{Jackson2019}
{Jackson}, B., {Adams}, E., {Sandidge}, W., {Kreyche}, S., \& {Briggs}, J.
  2019, \aj, 157, 239, \dodoi{10.3847/1538-3881/ab1b30}

\bibitem[{{Kataria} {et~al.}(2016){Kataria}, {Sing}, {Lewis}, {Visscher},
  {Showman}, {Fortney}, \& {Marley}}]{Kataria2016}
{Kataria}, T., {Sing}, D.~K., {Lewis}, N.~K., {et~al.} 2016, \apj, 821, 9,
  \dodoi{10.3847/0004-637X/821/1/9}

\bibitem[{{Keating} {et~al.}(2019){Keating}, {Cowan}, \& {Dang}}]{Keating2019}
{Keating}, D., {Cowan}, N.~B., \& {Dang}, L. 2019, Nature Astronomy, 3, 1092,
  \dodoi{10.1038/s41550-019-0859-z}

\bibitem[{{Komacek} \& {Showman}(2016)}]{Komacek2016}
{Komacek}, T.~D., \& {Showman}, A.~P. 2016, \apj, 821, 16,
  \dodoi{10.3847/0004-637X/821/1/16}

\bibitem[{{Komacek} {et~al.}(2017){Komacek}, {Showman}, \& {Tan}}]{Komacek2017}
{Komacek}, T.~D., {Showman}, A.~P., \& {Tan}, X. 2017, \apj, 835, 198,
  \dodoi{10.3847/1538-4357/835/2/198}

\bibitem[{{Lodders}(2010)}]{Lodders2010}
{Lodders}, K. 2010, Astrophysics and Space Science Proceedings, 16, 379,
  \dodoi{10.1007/978-3-642-10352-0_8}

\bibitem[{{Lorenz}(1963)}]{Lorenz1963}
{Lorenz}, E.~N. 1963, Journal of the Atmospheric Sciences, 20, 130,
  \dodoi{10.1175/1520-0469(1963)020<0130:DNF>2.0.CO;2}

\bibitem[{{Menou}(2012{\natexlab{a}})}]{Menou2012b}
{Menou}, K. 2012{\natexlab{a}}, \apjl, 754, L9,
  \dodoi{10.1088/2041-8205/754/1/L9}

\bibitem[{{Menou}(2012{\natexlab{b}})}]{Menou2012a}
---. 2012{\natexlab{b}}, \apj, 745, 138, \dodoi{10.1088/0004-637X/745/2/138}

\bibitem[{{Newcomb}(1961)}]{Newcomb1961}
{Newcomb}, W.~A. 1961, Physics of Fluids, 4, 391, \dodoi{10.1063/1.1706342}

\bibitem[{{Parker}(1975)}]{Parker1975}
{Parker}, E.~N. 1975, \apj, 198, 205, \dodoi{10.1086/153593}

\bibitem[{{Perna} {et~al.}(2010){Perna}, {Menou}, \& {Rauscher}}]{Perna2010a}
{Perna}, R., {Menou}, K., \& {Rauscher}, E. 2010, \apj, 719, 1421,
  \dodoi{10.1088/0004-637X/719/2/1421}

\bibitem[{{Price} {et~al.}(2012){Price}, {Link}, {Epstein}, \&
  {Li}}]{Price2012}
{Price}, S., {Link}, B., {Epstein}, R.~I., \& {Li}, H. 2012, \mnras, 420, 949,
  \dodoi{10.1111/j.1365-2966.2011.19807.x}

\bibitem[{{Rauscher} \& {Menou}(2010)}]{Rauscher2010}
{Rauscher}, E., \& {Menou}, K. 2010, \apj, 714, 1334,
  \dodoi{10.1088/0004-637X/714/2/1334}

\bibitem[{{Rauscher} \& {Menou}(2013)}]{Rauscher2013}
---. 2013, \apj, 764, 103, \dodoi{10.1088/0004-637X/764/1/103}

\bibitem[{{Read} \& {Lebonnois}(2018)}]{Read2018}
{Read}, P.~L., \& {Lebonnois}, S. 2018, Annual Review of Earth and Planetary
  Sciences, 46, 175, \dodoi{10.1146/annurev-earth-082517-010137}

\bibitem[{{Rogers}(2017)}]{Rogers2017b}
{Rogers}, T.~M. 2017, Nature Astronomy, 1, 0131,
  \dodoi{10.1038/s41550-017-0131}

\bibitem[{{Rogers} \& {Komacek}(2014)}]{Rogers2014b}
{Rogers}, T.~M., \& {Komacek}, T.~D. 2014, \apj, 794, 132,
  \dodoi{10.1088/0004-637X/794/2/132}

\bibitem[{{Rogers} \& {McElwaine}(2017)}]{Rogers2017a}
{Rogers}, T.~M., \& {McElwaine}, J.~N. 2017, \apjl, 841, L26,
  \dodoi{10.3847/2041-8213/aa72da}

\bibitem[{{Showman} {et~al.}(2009){Showman}, {Fortney}, {Lian}, {Marley},
  {Freedman}, {Knutson}, \& {Charbonneau}}]{Showman2009}
{Showman}, A.~P., {Fortney}, J.~J., {Lian}, Y., {et~al.} 2009, \apj, 699, 564,
  \dodoi{10.1088/0004-637X/699/1/564}

\bibitem[{{Showman} \& {Guillot}(2002)}]{Showman2002}
{Showman}, A.~P., \& {Guillot}, T. 2002, \aap, 385, 166,
  \dodoi{10.1051/0004-6361:20020101}

\bibitem[{{Showman} \& {Polvani}(2011)}]{Showman2011}
{Showman}, A.~P., \& {Polvani}, L.~M. 2011, \apj, 738, 71,
  \dodoi{10.1088/0004-637X/738/1/71}

\bibitem[{{Tritton}(1988)}]{Tritton1988}
{Tritton}, D.~J. 1988, {Physical fluid dynamics /2nd revised and enlarged
  edition/} (Osford Science), 246--254

\bibitem[{{von Essen} {et~al.}(2020){von Essen}, {Mallonn}, {Borre}, {Antoci},
  {Stassun}, {Khalafinejad}, \& {Tautvai{\v{s}}ien{\.{e}}}}]{vonEssen2020}
{von Essen}, C., {Mallonn}, M., {Borre}, C.~C., {et~al.} 2020, \aap, 639, A34,
  \dodoi{10.1051/0004-6361/202037905}

\bibitem[{{Yadav} \& {Thorngren}(2017)}]{Yadav2017}
{Yadav}, R.~K., \& {Thorngren}, D.~P. 2017, \apjl, 849, L12,
  \dodoi{10.3847/2041-8213/aa93fd}

\end{thebibliography}
\bibliographystyle{aasjournal}
\end{document}